\documentclass{svjour3}                     
\smartqed  
\usepackage{graphicx}
\journalname{Few-Body Systems (FB20)}
\begin{document}

\title{Effective interactions in the delta-shells potential
\thanks{Supported by Spanish DGI
  (grant FIS2011-24149) and Junta de Andaluc{\'{\i}a} (grant FQM225).
  R.N.P. is supported by a Mexican CONACYT grant.}  
\thanks{Presented by R.N.P. at the 20th International
IUPAP Conference on Few-Body Problems in Physics, 20 - 25 August,
  2012, Fukuoka, Japan} }
\author{R. Navarro P\'erez       \and J. E. Amaro \and \\  
E. Ruiz Arriola}


\institute{R. Navarro P\'erez \at
Departamento de F\'{\i}sica At\'{o}mica, Molecular y
  Nuclear and Instituto Carlos I de F{\'\i}sica Te\'orica y Computacional 
Universidad de Granada, E-18071 Granada, Spain. \\
              \email{rnavarrop@ugr.es}           
}

\date{Received: 27-IX-2012}

\maketitle

\begin{abstract}
We determine two-body Skyrme force parameters from a Nucleon-Nucleon
interaction as a function of the maximal momentum fitting NN
scattering data. We find general agreement with $V_{\rm low k}$
interactions based on high quality potentials. 
\end{abstract}


\newcommand{\bea}{\begin{eqnarray}}
\newcommand{\eea}{\end{eqnarray}}

\vskip1cm 

The use of effective interactions in Nuclear Physics is rather old and
dates back to the pioneering works of Moshinsky~\cite{Moshinsky195819}
and Skyrme~\cite{Skyrme:1959zz}. One of the advantages in doing so is
that, as compared to {\it ab initio} calculations, the nuclear many
body wave function has a much simpler structure since short range
correlations play a marginal role allowing for a fruitful
implementation of mean field Hartree-Fock
calculations~\cite{Vautherin:1971aw,Chabanat:1997qh,Bender:2003jk}.

At the two body level the effective interaction of
Moshinsky~\cite{Moshinsky195819} and Skyrme~\cite{Skyrme:1959zz} reads 
\begin{eqnarray} 
 V ({\bf
    p}',{\bf p}) 
&=& \int d^3 x e^{-i {\bf x}\cdot ({\bf p'}-{\bf p})}  \hat V({\bf x} ) 
 \nonumber \\ &=&  t_0 (1 + x_0 P_\sigma ) + \frac{t_1}2(1 + x_1
  P_\sigma ) ({\bf p}'^2 + {\bf p}^2) \nonumber \\ &+&  
 t_2 (1 + x_2
  P_\sigma ) {\bf p}' \cdot {\bf p} + 2 i W_0 {\bf S} \cdot({\bf p}'
  \wedge {\bf p}) \nonumber \\ &+& 
\frac{t_T}2 \left[ \sigma_1 \cdot {\bf p}
  \, \sigma_2 \cdot {\bf p}+ \sigma_1 \cdot {\bf p'} \, \sigma_2
  \cdot {\bf p'} - \frac13 \sigma_1 \, \cdot 
\sigma_2 ({\bf p'}^2+  {\bf p}^2)
\right]  \nonumber \\ &+& 
\frac{t_U}2 \left[ \sigma_1 \cdot {\bf p}
  \, \sigma_2 \cdot {\bf p}'+ \sigma_1 \cdot {\bf p'} \, \sigma_2
  \cdot {\bf p} - \frac23 \sigma_1 \, \cdot 
\sigma_2 {\bf p'}\cdot  {\bf p}
\right]  
+ {\cal O} (p^4) 
\label{eq:skyrme2}
\end{eqnarray} 
where $P_\sigma = (1+ \sigma_1 \cdot \sigma_2)/2$ is the spin exchange
operator with $P_\sigma=-1$ for spin singlet $S=0$ and $P_\sigma=1$
for spin triplet $S=1$ states. In Ref.~\cite{Arriola:2010hj} the
parameters of Eq.~(\ref{eq:skyrme2}) where determined from just NN
threshold properties such as scattering lengths, effective ranges and
volumes without explicitly taking into account the finite range of the
NN interaction.

Our aim here is to compute the Skyrme parameters from an analysis of
NN scattering data below pion production threshold rather than to a
fit to double-closed shell nuclei and nuclear matter saturation
properties as it is usually
done~\cite{Vautherin:1971aw,Chabanat:1997qh,Bender:2003jk}.

While the idea of effective interactions is conceptually simple and
rather appealing computationally there is no unique or universal
definition since any different method assumes a particular off-shell
extrapolation which cannot be tested experimentally. On the other hand
effective interactions such as Eq.~(\ref{eq:skyrme2}) truly 
depend on the relevant wavelengths involved.
This becomes clear in the $V_{\rm low k}$
method~\cite{Bogner:2003wn,Bogner:2009bt}, where one truncates the
model space Hamiltonian for states with CM momentum $p \le \Lambda$.
However, in order to be able to implement this method one needs a {\it
  choice} of a phenomenological potential which besides fitting the
data satisfactorily provides the half-off-shell scattering amplitude.

In the present contribution we define the interaction
Eq.~(\ref{eq:skyrme2}) by fitting a potential to NN phase-shifts below
a given maximum CM momentum $p \le \Lambda$. To proceed we use the 
delta-shells local potential  for partial waves
\begin{eqnarray}
V_{l,l'}^{JS} (r) =  \sum_{i} (\lambda_{l,l'}^{J,S}) \delta (r-r_i)  \qquad r \le 3 {\rm fm}
\label{eq:ds-pot}
\end{eqnarray}
proposed by Avil\'es long ago~\cite{Aviles:1973ee} for the short range
part. In addition we take the One-Pion-Exchange (OPE) and
electromagnetic tail for $r \ge 3 {\rm fm}$.  With this potential and
below pion production threshold $p \le \sqrt{M_N m_\pi} = 362 {\rm
  MeV}$ a description of scattering observables can be achieved with
$\chi^2/{\rm d.o.f} =1.06$~\cite{Perez:2012qf}, of comparable high
quality as the bench-marking fits of the Nijmegen
group~\cite{Stoks:1993tb,Stoks:1994wp} and subsequent
AV18~\cite{Wiringa:1994wb}, CD-Bonn~\cite{Machleidt:2000ge} and
Spectator model~\cite{Gross:2008ps} potentials. The rationale behind
the schematic form of Eq.~(\ref{eq:ds-pot}) is based on the
expectation that a coarse graining of the interaction to a given
wavelength should not display fluctuations of the interactions to
shorter distances than $\Delta r \sim 1/\sqrt{M_N m_\pi}=0.54 {\rm
  fm}$.

\begin{figure*}
\includegraphics[width=4cm,height=4cm]{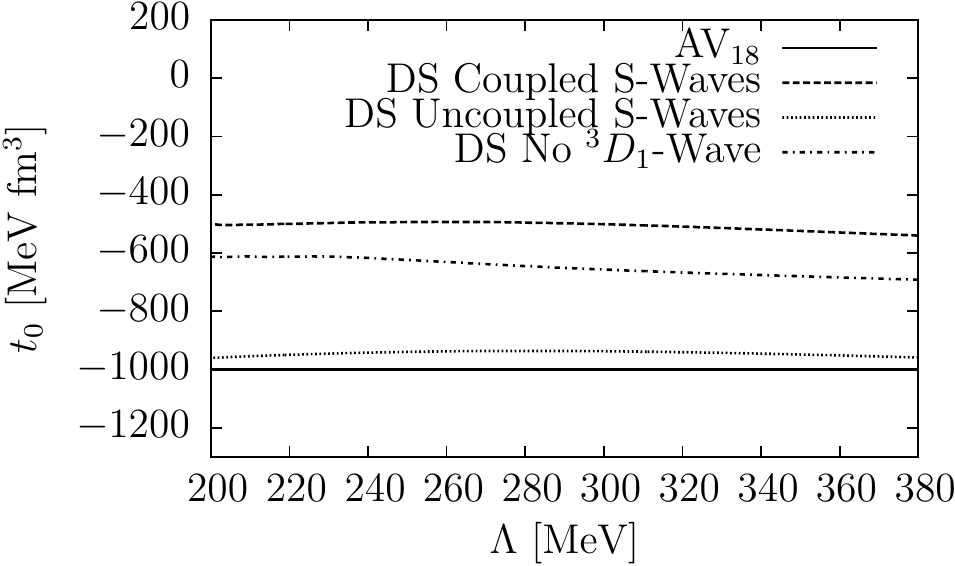}
\includegraphics[width=4cm,height=4cm]{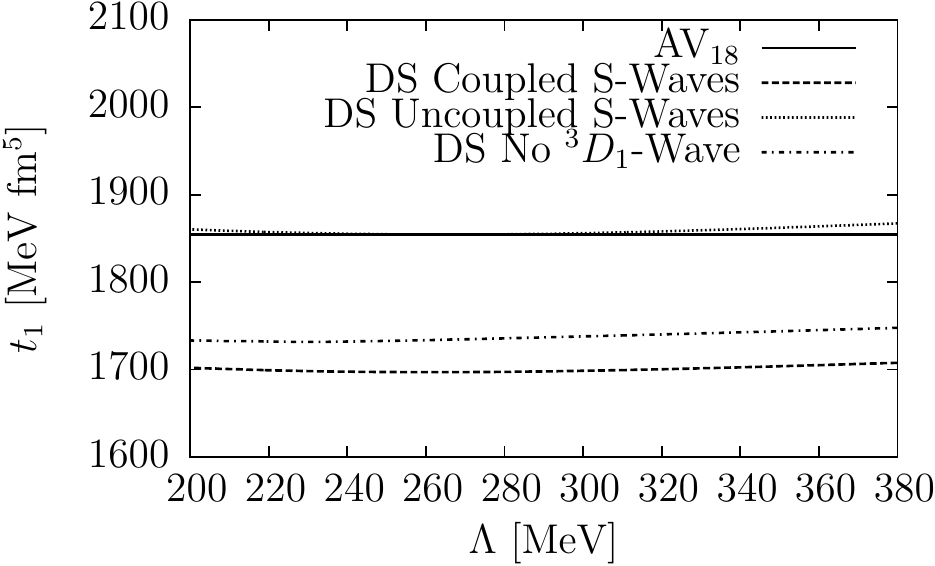}
\includegraphics[width=4cm,height=4cm]{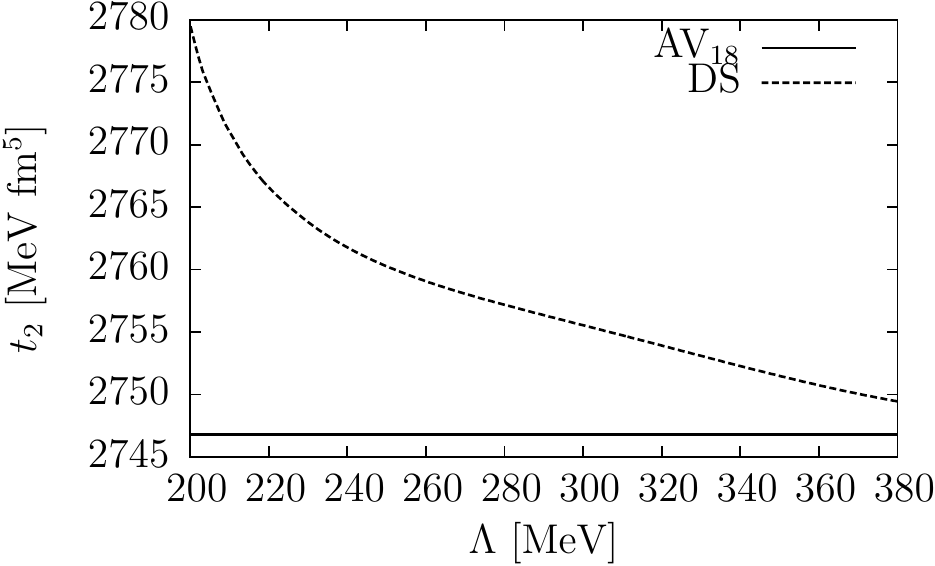}
\includegraphics[width=4cm,height=4cm]{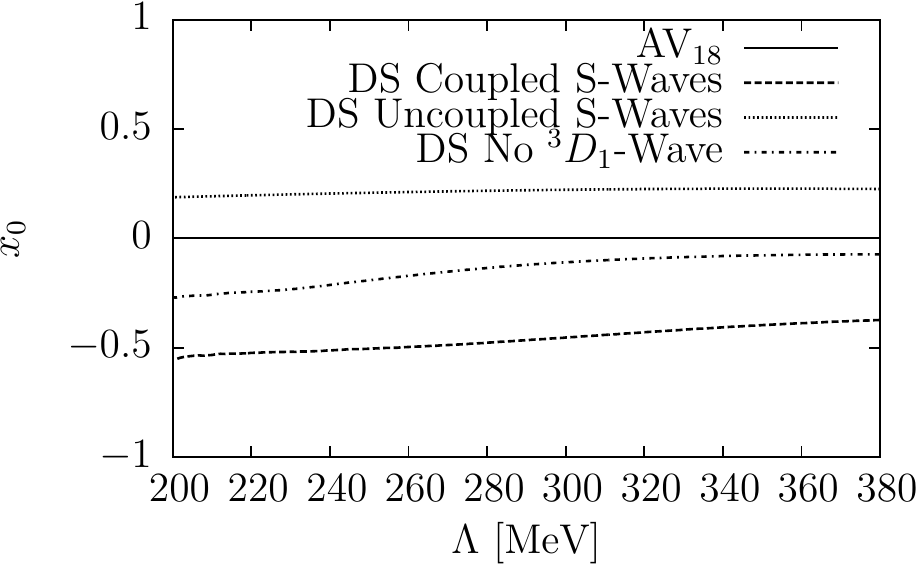}
\includegraphics[width=4cm,height=4cm]{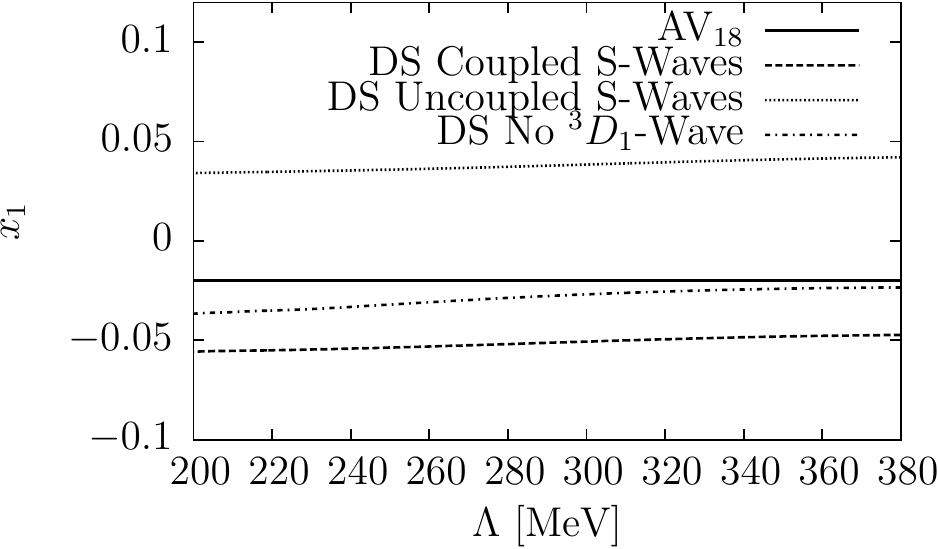}
\includegraphics[width=4cm,height=4cm]{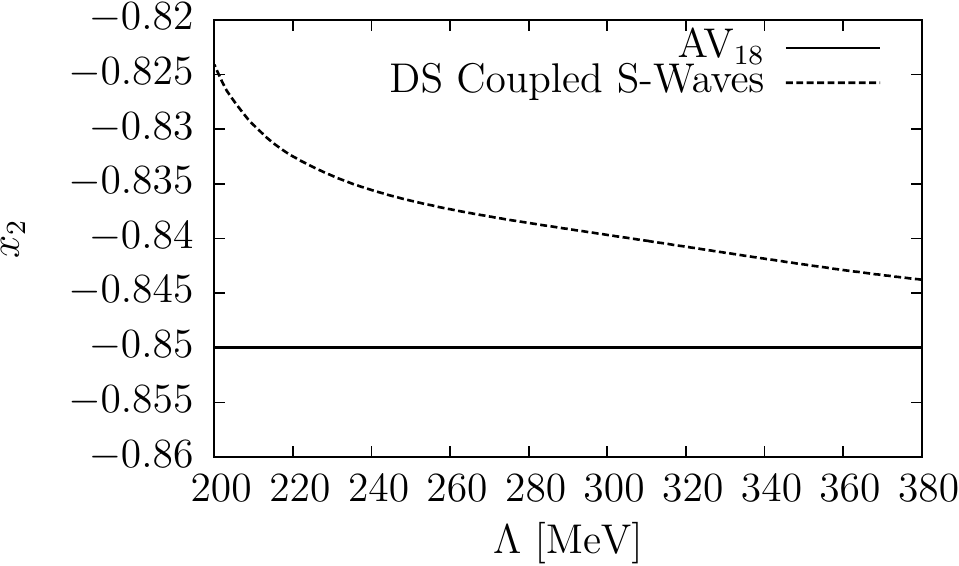}
\includegraphics[width=4cm,height=4cm]{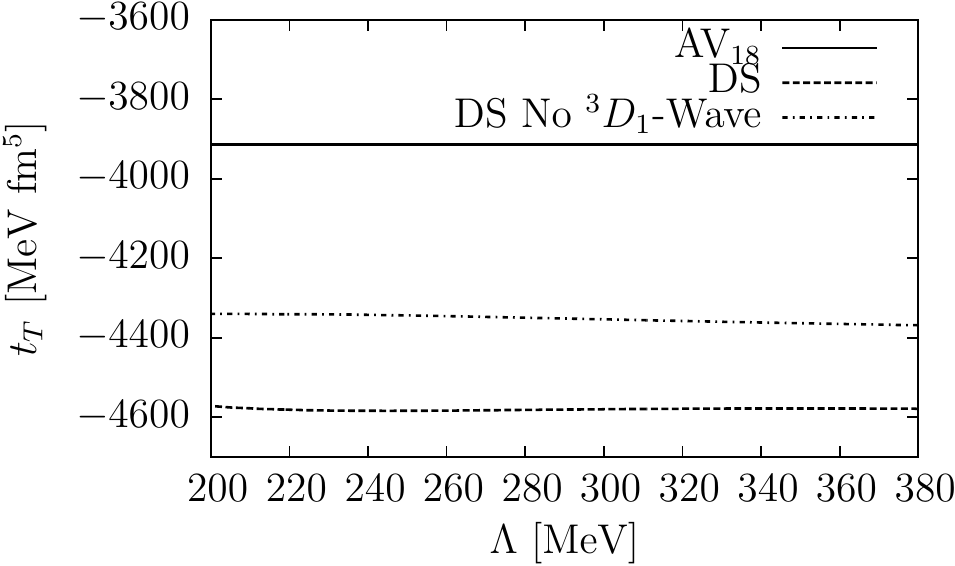}
\includegraphics[width=4cm,height=4cm]{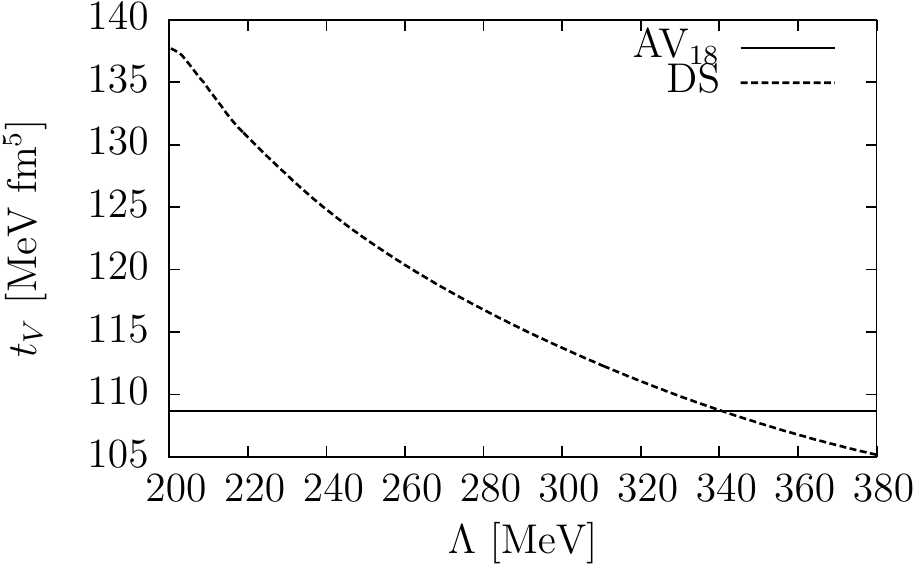}
\includegraphics[width=4cm,height=4cm]{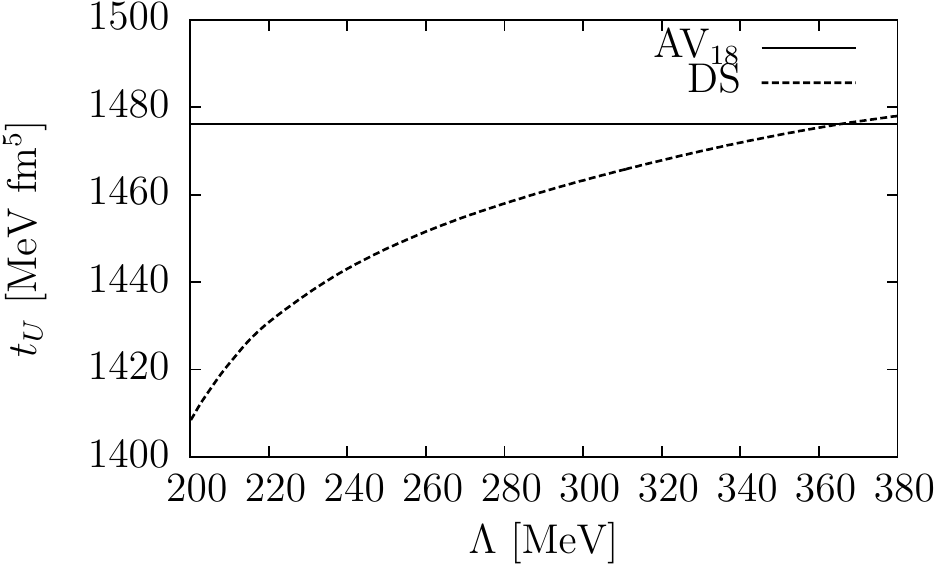}
\caption{Dependence of the effective interaction parameters as a
  functions of the maximal CM momentum $\Lambda$ (in MeV) for which
  the neutron-protron phase shifts have been fitted. We compare with
  the AV18-$V_{\rm low k}$ potential when $\Lambda_{\rm low k}=2.1 {\rm fm}^{-1}$}
\label{fig:2}       
\end{figure*}

For the  potential in 
Eq.~(\ref{eq:ds-pot}) one finds after some calculation the result  
\bea
(t_0, x_0 t_0) &=& \frac12 \int\, d^3 x\,  \,  \left[ V_{^3S_1}(r) \pm V_{^1S_0}(r) \right] 
\, , \nonumber \\  
(t_1, x_1 t_1)  &=& -\frac1{12} \int\, d^3 x\, r^2 \,  \left[ V_{^3S_1}(r) \pm V_{^1S_0}(r) \right] 
\, , \nonumber \\  
(t_2, x_2 t_2)  &=& \frac{1}{54} \int \, d^3 x\, r^2 \, \left[
  V_{^3P_0}(r) + 3 V_{^3P_1}(r) + 5 V_{^3P_2}(r)\pm 9 V_{^1P_1}(r)
  \right] \, , \nonumber \\   
t_V= W_0 &=& \frac{1}{72} \int \, d^3 x\,  r^2 \, \left[ 2 V_{^3P_0} (r) +  3 V_{^3P_1} (r) -5  V_{^3P_2} (r) \right] \, , \nonumber \\  
t_T &=& \frac{1}{5 \sqrt{2}} \int \, d^3 x\, r^2 \,  V_{E_1}(r) \, , \nonumber \\t_U &=& \frac{1}{36} \int \, d^3 x\, r^2 \, \left[
-2  V_{^3P_0}(r) + 3 V_{^3P_1}(r) - V_{^3P_2}(r)\right] \, , 
\label{eq:skyrme}
\eea where the $\pm$ in the first three equations refers to the first
and second possibilities on the l.h.s. We use the delta-shells
potential for $r \le r_c$.  The effective interaction due to OPE above
$r > r_c$ is given by the following formulas in the isospin invariant
case
\begin{eqnarray}
t_0 |_{\rm  OPE} &=& -\frac{f_{\pi NN}^2 }{m_\pi^2} \Gamma (2,m_\pi r_c) 
\, ,\qquad 
x_0 t_0 |_{\rm  OPE}= 0 
\, ,\nonumber \\
t_1 |_{\rm  OPE} &=&  \frac{f_{\pi NN}^2 }{3 m_\pi^4} \Gamma (4,m_\pi r_c) 
\, ,\qquad 
x_1 t_1 |_{\rm  OPE} = 0 
\, ,\nonumber \\
t_2 |_{\rm  OPE} &=&  \frac{5 f_{\pi NN}^2 }{9 m_\pi^4} \Gamma (4,m_\pi r_c) 
\, ,\qquad 
x_2 t_2 |_{\rm  OPE} =  -\frac{4 f_{\pi NN}^2 }{9 m_\pi^4} \Gamma (4,m_\pi r_c)
\, ,\nonumber \\
t_V |_{\rm  OPE} &=&  0 
\, ,\nonumber \\
t_U |_{\rm  OPE} &=&  \frac{2 f_{\pi NN}^2 }{15 m_\pi^4}
\left[3 \Gamma (2,m_\pi r_c)+ 
3 \Gamma (3,m_\pi r_c)+
\Gamma (4,m_\pi r_c) \right] 
\, ,\nonumber \\
t_T |_{\rm  OPE} &=&  -\frac{2 f_{\pi NN}^2 }{5 m_\pi^4}
\left[3 \Gamma (2,m_\pi r_c)+ 
3 \Gamma (3,m_\pi r_c)+
\Gamma (4,m_\pi r_c) \right] 
\, , 
\end{eqnarray}
where $f_{\pi NN} = g_{\pi NN} m_\pi /2 M_N $, with $f_{\pi
  NN}^2/(4\pi) \sim 0.08$ and $\Gamma (n, x) = \int_x^\infty \, dt \,
e^{-t} t^{n-1} $. These OPE contributions are numerically tiny 
for $r_c= 3 {\rm fm}$. From the fit in Ref.~\cite{Perez:2012qf} we
get the results of Table~\ref{tab:par}. They are compared with the extraction 
of Ref.~\cite{Arriola:2010hj}. 


The scale dependence on the fitted np scattering phase-shifts is
presented in Fig.~\ref{fig:2}. Note that the set of
Eqs.~(\ref{eq:skyrme}) involve only S- and P-waves as well as the
SD-wave mixing but no D-waves. However, the tensor force requires a
non-vanishing D-wave.  Having this in mind we distinguish three
different situations for the case of the triplet $^3S_1$ wave due to
the role played by the tensor force. The ``coupled case'' corresponds
to make a fit to the $^3S_1,^3D_1, E_1$ phase-shifts, whereas the
``uncoupled case'' is obtained from a fit of the $^1S_3$ potential
from the $^3S_1$ phase shift, without considering the coupling to the
$^3D_1$ channel. Another intermediate case corresponds to just fit
$^3S_1$ and $E_1$ phases taking a vanishing $^3D_1$ potential. As can
be seen from Fig.~\ref{fig:2} the uncoupled case resembles best the
$V_{\rm low k}$-value.  Generally, there is a close agreement (note
the y-axis scales) with the $V_{\rm low k}$ results as applied to the
AV18 potential when a value of $\Lambda_{\rm low k}=2.1 {\rm fm}^{-1}$ is
taken. A more complete analysis properly weighting the relative
importance of $D-$ vs $P-$ and $S-$waves with inclusion of
uncertainties~\cite{NavarroPerez:2012vr,Perez:2012kt} will be
presented elsewhere. In any case, the enhanced attraction confirms the
binding features of doubled closed shell nuclei outlined in
Ref.~\cite{NavarroPerez:2011fm}.

%
%
\begin{table}
\caption{ Skyrme parameters obtained from the AV18-$V_{\rm low k}$
  with $\Lambda_{\rm low k}=2.1 {\rm fm}^{-1}$ and the charge dependent
  Delta-shells fit up to $E_{\rm LAB}=350 {\rm MeV}$ . Both potentials
  reproduce the np phase-shifts up to pion production
  threshold. $t_{0,1,2}$ are in $({\rm MeV}{\rm fm}^3)$, $x_{0,1,2}$
  are dimensionless and $t_{U,V,W}$ are in $({\rm MeV}{\rm fm}^5)$.}
\label{tab:par} 
\begin{tabular}{lllllllllll}
\hline\noalign{\smallskip}
Parameter  & $t_0 $  &  $x_0 $ & $t_1 $  & $x_1 $ &  $t_2 $ & $x_2$ & $W_0$ & $t_U $ & $t_T $ \\
\noalign{\smallskip}\hline\noalign{\smallskip}
$V_{\rm low k}$  &   -999.6   &   0.002 & 1854.2  & -0.02 & 2198.3 &
-0.91  & 84.1 & 1235.2  &  -3864.0 \\ 
Delta-shell   &   -555.3 &  -0.36 & 1711.8 & -0.05 & 2746.8 & 
-0.845 &  108.7 &  1476.2 &-4576.4 \\  
\noalign{\smallskip}\hline
\end{tabular}
\end{table}



\end{document}